\documentclass[useAMS,usenatbib,usegraphicx]{mn2e}

\title[A double main sequence turn-off in NGC 1846]{A double main sequence turn-off in the 
rich star cluster NGC 1846 in the Large Magellanic Cloud}
\author[A.~D.~Mackey \& P.~B.~Nielsen]{A.~D.~Mackey$^{1}$\ and P.~Broby~Nielsen$^{1}$\\
$^{1}$Institute for Astronomy, University of Edinburgh, Royal Observatory, Blackford Hill, 
Edinburgh, EH9 3HJ, UK}

\begin{document}

\date{Draft version \today}

\pagerange{\pageref{firstpage}--\pageref{lastpage}} \pubyear{2007}

\maketitle

\label{firstpage}

\begin{abstract}
We report on HST/ACS photometry of the rich intermediate-age star cluster NGC 1846 in
the Large Magellanic Cloud, which clearly reveals the presence of a double main 
sequence turn-off in this object. Despite this, the main sequence, sub-giant branch,
and red giant branch are all narrow and well-defined, and the red clump is compact.
We examine the spatial distribution of turn-off stars and demonstrate that all
belong to NGC 1846 rather than to any field star population. In addition, the spatial 
distributions of the two sets of turn-off stars may exhibit different central concentrations 
and some asymmetries. By fitting isochrones, we show that the properties of the 
colour-magnitude diagram can be explained if there are two stellar populations of equivalent 
metal abundance in NGC 1846, differing in age by $\approx 300$ Myr. The absolute ages 
of the two populations are $\sim 1.9$ and $\sim 2.2$ Gyr, although there may be a 
systematic error of up to $\pm 0.4$ Gyr in these values. The metal abundance inferred
from isochrone fitting is $[$M$/$H$] \approx -0.40$, consistent with spectroscopic 
measurements of $[$Fe$/$H$]$. We propose that the observed properties of NGC 1846 can 
be explained if this object originated via the tidal capture of two star clusters 
formed separately in a star cluster group in a single giant molecular cloud. This 
scenario accounts naturally for the age difference and uniform metallicity 
of the two member populations, as well as the differences in their spatial distributions.
\end{abstract}

\begin{keywords}
globular clusters: individual: NGC 1846 -- Magellanic Clouds -- galaxies: star clusters.
\vspace{-4mm}
\end{keywords}

\section{Introduction}
The Large Magellanic Cloud (LMC) possesses an extensive system of massive star clusters,
covering the full range of ages $10^6 - 10^{10}$ yr. This age spread, which is not
present in the Galactic globular cluster population, combined with the close
proximity of the LMC system, means that these objects have proved vital to our understanding
of star cluster formation and evolution. Furthermore, the age and metallicity
distributions of the LMC clusters, together with their kinematics, offer important 
insights into the formation and subsequent development of the LMC itself.

As part of a study of the structural evolution of massive stellar clusters, we have
conducted a snapshot imaging survey of some $\sim 50$ such objects in the LMC and SMC
using the Advanced Camera for Surveys (ACS) on-board the Hubble Space Telescope (HST).
Many of the target clusters have not previously been investigated in any 
detail, and we are therefore working on deriving accurate photometric
ages and metallicities for the sample \citep*{mackey:04,mackey:06}. While reducing
our observations of the poorly-studied rich intermediate-age LMC cluster NGC 1846, 
we recently noticed that the colour-magnitude diagram (CMD) for this object exhibited 
a very peculiar main sequence turn-off region; further scrutiny revealed clearly 
the presence of two distinct turn-offs. With a few notable exceptions (for example,
the massive Galactic globular cluster $\omega$ Centauri), the vast majority of star 
clusters which have been resolved by observations into individual stars are comprised
of single stellar populations -- that is, groups of stars formed at the same time and 
with the same chemical composition. NGC 1846 is therefore an important object, as
it can potentially offer us new insights into the processes of star and star cluster
formation.

In this paper we present our photometry of NGC 1846 together with a detailed 
investigation of its extremely unusual CMD (Sections \ref{s:data} and \ref{ss:cmd}). 
We examine the spatial distribution of stars in the main sequence turn-off region, and 
demonstrate that both turn-offs are clearly associated with the cluster rather than 
being an artifact introduced by field star contamination (Section \ref{ss:spatial}). 
In addition, we show that neither differential reddening, nor a significant 
line-of-sight depth to NGC 1846 can explain the double turn-off. From isochrone fitting, 
we determine the ages and metallicities of the two populations present in NGC 1846 
(Section \ref{ss:isochrone}), and use these results to constrain the possible means by 
which this object can have been formed (Section \ref{s:discussion}).

\begin{figure*}
\centering
\includegraphics[width=165mm]{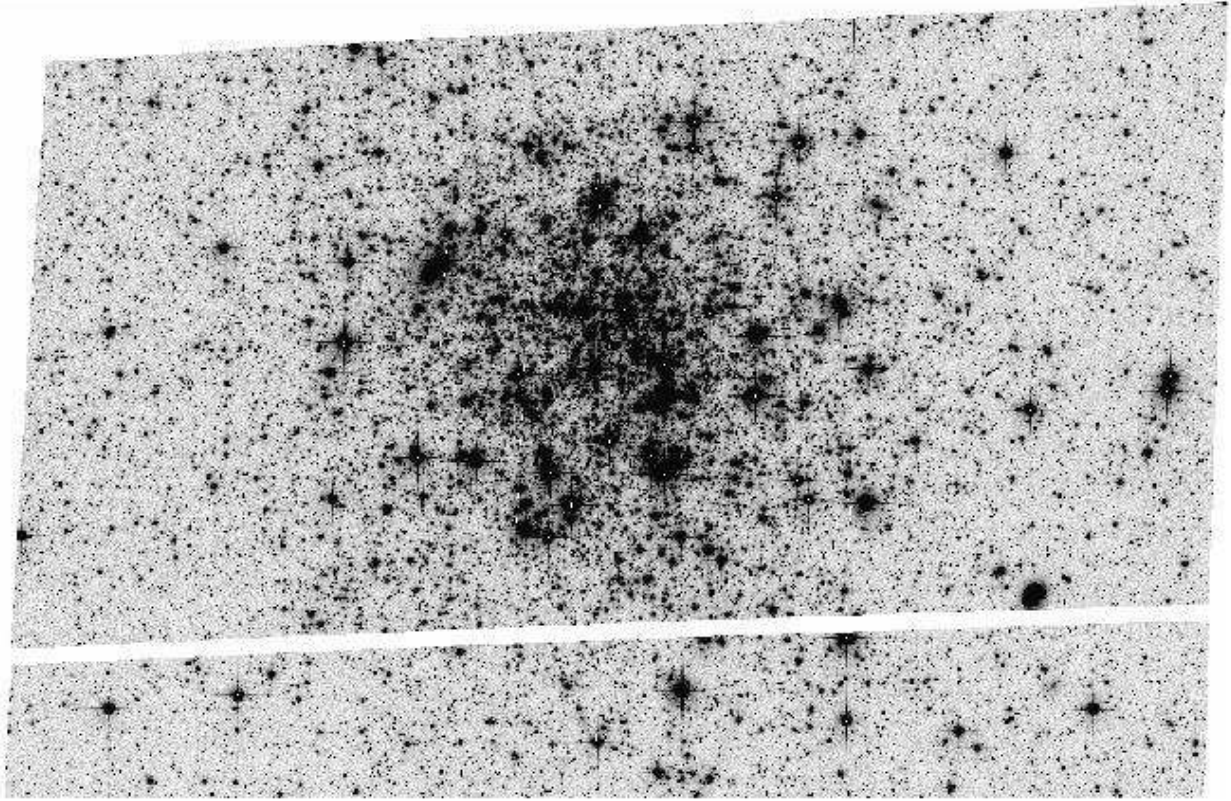}
\vspace{3mm}
\caption{Drizzled ACS/WFC F814W image of the core of NGC 1846 (exposure $200$s). The full 
image has been cropped to include only WFC chip 1 and part of WFC chip 2 -- these cover 
the central cluster region. Note that the brightest stars in the cluster are saturated.}
\label{f:image}
\end{figure*}

\section{Observations and data reduction}
\label{s:data}
Our observations were made during HST Cycle $12$ on 2003 October 08 using the ACS 
Wide Field Channel (WFC).
As a snapshot target, one frame was taken in each of two filters -- F555W (dataset 
j8ne55z9q) and F814W (dataset j8ne55zeq). Exposure durations were $300$s and $200$s, 
respectively. The ACS WFC consists of two $2048\times 4096$ pixel CCDs separated by a 
gap $\approx 50$ pixels wide. The plate scale is $0.05$ arcsec per pixel, resulting in 
a total areal coverage of approximately $202\times 202$ arcsec. The core of NGC 
1846 was positioned at the centre of chip 1 so that the inter-chip gap did not impact 
on the innermost region of the cluster. In order to help with the identification and 
removal of hot-pixels and cosmic rays, the F814W image was offset from the F555W image 
by $\approx 2$ pixels in both the $x$- and $y$-directions. 

The data products produced by the STScI reduction pipeline, which we retrieved via the
public archive, have had bias and dark-current frames subtracted and are divided
by a flat-field image. In addition, known hot-pixels and other defects are masked,
and the photometric keywords in the image headers are calculated. We also obtained
distortion-corrected (drizzled) images from the archive, produced using the {\sc pyraf} 
task {\sc multidrizzle}. Part of the drizzled F814W image is displayed in Fig.
\ref{f:image}.

We used the {\sc dolphot} photometry software \citep{dolphin:00}, specifically the ACS
module, to photometer our flatfielded F555W and F814W images. {\sc dolphot} performs 
point-spread function fitting using PSFs especially tailored to the ACS camera. Before 
performing the photometry, we first prepared the images using the {\sc dolphot} packages 
{\sc acsmask} and {\sc splitgroups}. Respectively, these two packages apply the image 
defect mask and then split the multi-image STScI FITS files into a single FITS file per 
chip. We then used the main {\sc dolphot} routine to simultaneously make photometric
measurements on the pre-processed images, relative to the coordinate 
system of the drizzled F814W image. We chose to fit the sky locally around each
detected source (important due to the crowded nature of the target), and keep only
objects with a signal greater than $10$ times the standard deviation of the background.
The output photometry from {\sc dolphot} is on the calibrated VEGAMAG scale of
\citet{sirianni:05}, and corrected for charge-transfer efficiency (CTE) degradation.

To obtain a clean list of stellar detections with high quality photometry, we applied
a filter employing the sharpness and ``crowding'' parameters calculated by {\sc dolphot}.
The sharpness is a measure of the broadness of a detected object relative to the
PSF -- for a perfectly-fit star this parameter is zero, while it is negative for an
object which is too sharp (perhaps a cosmic-ray) and positive for an object which is
too broad (say, a background galaxy). The crowding parameter measures how much brighter
a detected object would have been measured had nearby objects not been fit simultaneously.
We selected only objects with $-0.15 \leq {\rm sharpness} \leq 0.15$ in both frames, and 
${\rm crowding} \leq 0.5$ mag in both frames. We also only kept objects classified by
{\sc dolphot} as good stars (object type $1$), as opposed to elongated or extended objects
(object types $> 1$).

\section{Results}
\label{s:results}
\subsection{Colour-magnitude diagram}
\label{ss:cmd}
\begin{figure}
\centering
\includegraphics[width=0.49\textwidth]{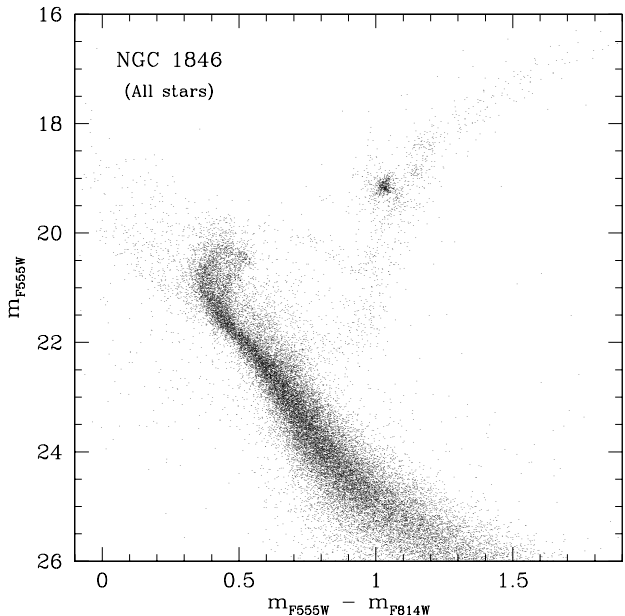}
\caption{Colour-magnitude diagram for NGC 1846. All detected sources which passed
successfully through our quality filter are plotted ($34534$ objects). The main cluster
sequences are clearly visible; however there is some field star contamination present
in key regions, particularly along the RGB and main sequence. Even so, the double
main sequence turn-off is strikingly evident.\vspace{-1mm}}
\label{f:cmdall}
\end{figure}

\begin{figure}
\centering
\includegraphics[width=0.49\textwidth]{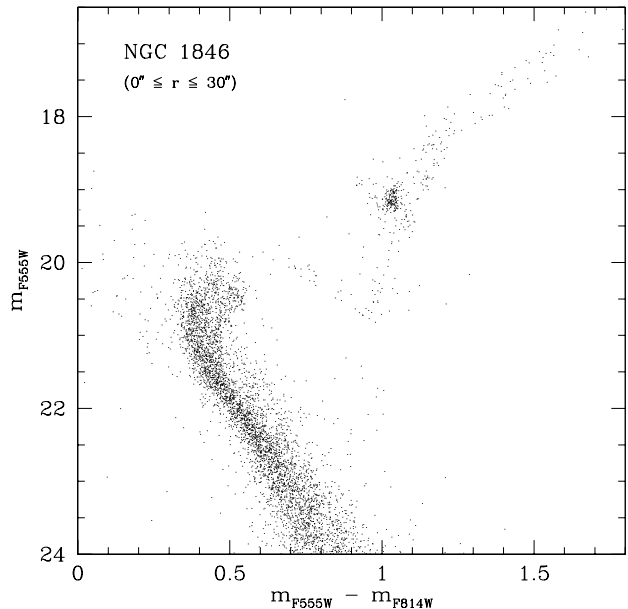}
\caption{Colour-magnitude diagram for NGC 1846, now plotted using only quality-filtered
stars at projected radial distances less than $30\arcsec$ from the centre ($6599$ objects). 
The majority of field star contamination has been removed leaving the cluster
sequences clearly visible, particularly the double main sequence turn-off. Note that 
the view in this plot is focussed on the turn-off region, SGB and RGB, so the
scale is different from that of Fig. \ref{f:cmdall}.\vspace{-1mm}}
\label{f:cmdrad}
\end{figure}

The colour-magnitude diagram (CMD) for NGC 1846 is presented in Fig. \ref{f:cmdall}. 
All detected stars which passed successfully through the quality filter are plotted. 
There is some evident field-star contamination present in this CMD, so in Fig. 
\ref{f:cmdrad} we plot only stars within $30\arcsec$ of the cluster centre, which
lies near pixel coordinates $(x_c,y_c) = (2120,3150)$ (see Section \ref{ss:spatial} below).
Since this selected region is very much smaller in area than the full ACS coverage, field-star
contamination in the CMD is greatly reduced compared to Fig. \ref{f:cmdall} and allows 
one to clearly identify the primary cluster sequences. In principle we could apply a 
more sophisticated statistical subtraction of field stars; however this is unnecessary to 
achieve the goals of the present work. Selecting the innermost cluster regions, as in
Fig. \ref{f:cmdrad}, clearly offers a sufficiently uncontaminated CMD.

The most striking and unusual feature of the CMD for NGC 1846 is the presence of an 
apparently double main sequence turn-off. This is clearly visible in both Figs. 
\ref{f:cmdall} and \ref{f:cmdrad}, suggesting it is not an artifact of field star 
contamination (we address this issue more completely in Section \ref{ss:spatial} 
below). Apart from this, the CMD is as expected for an intermediate-age Magellanic 
Cloud cluster. There is a relatively narrow main sequence, a down-sloping sub-giant branch 
(SGB), and a narrow red giant branch (RGB), as well as a compact red clump (RC) near 
$m_{{\rm F555W}} \approx 19.1$ and $m_{{\rm F555W}}-m_{{\rm F814W}} \approx 1.0$. 
On the RGB just brighter than the RC level a clear knot of stars is visible, especially
in Figure \ref{f:cmdall}. This is possibly the RGB Bump -- 
the point at which the hydrogen-burning shell in a red giant star reaches the 
discontinuity left in the hydrogen abundance profile due to inner penetration of
convection. A similar feature has recently been observed in the rich intermediate-age 
LMC cluster NGC 1978 by \citet{mucciarelli:07} -- an object which presents a very
similar CMD to that for NGC 1846 (although without the double main sequence turn-off).

It is worth noting that a third turn-off is clearly visible in Fig. \ref{f:cmdall}, 
near $m_{{\rm F555W}} \sim 22.5$. This is clearly part of the contaminating
field star population (since it disappears in Fig. \ref{f:cmdrad}), and is
representative of a much older population than that present in NGC 1846.
\citet{mucciarelli:07} observe a similar population in the field surrounding
NGC 1978, and interpret it as the signature of a major star formation episode
which occurred $\sim 5-6$ Gyr ago when the LMC and SMC were gravitationally bound.

Above $m_{{\rm F555W}} \approx 18.0$, the width of the cluster RGB appears to increase
significantly. Examining the data quality flags produced by {\sc dolphot} we found 
that all stars above this level were flagged as being saturated in their central pixels
(see e.g., the brightest stars in Fig. \ref{f:image}). 
The accuracy of all photometry on the RGB above $m_{{\rm F555W}} = 18.0$ is therefore 
dubious, and the apparent increased spread is most likely artificial.

Based on an unpublished CMD, \citet{grocholski:06} suggested that NGC 1846 suffers 
from differential reddening;
however, from the narrow (lower) RGB and the compact RC visible in
Fig. \ref{f:cmdrad} we see no evidence for significant differential reddening, 
at least in the central regions of the cluster. Certainly the apparent double main
sequence turn-off cannot be produced by such an effect without introducing a large
spread into the other features on the cluster CMD. Similarly, the observed
lack of such a spread in these features militates against a large line-of-sight
depth in the system being responsible for the properties of the turn-off region.

\subsection{Spatial distribution of turn-off stars}
\label{ss:spatial}
The most intriguing aspect of the CMD for NGC 1846 is the double main sequence 
turn-off. Such a clearly defined example has not previously been observed in a
Magellanic Cloud cluster -- indeed, it is rare to find such a feature in {\it any}
type of star cluster. The turn-off region is suggestive of two separate cluster
populations. If this is so then it is important to ascertain whether there are any
differences in their spatial distributions, as is observed, for example, for the multiple 
populations found in the peculiar Galactic globular cluster $\omega$ Centauri 
\citep[see e.g.,][ and references therein]{villanova:07}.

\begin{figure}
\centering
\includegraphics[width=0.49\textwidth]{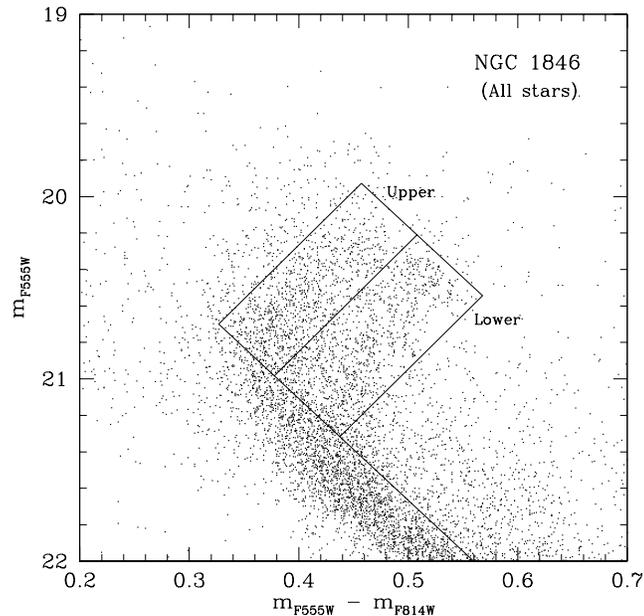}
\caption{Colour-magnitude diagram of the main sequence turn-off region of NGC 1846, 
with all stars plotted as in Fig. \ref{f:cmdall}. The two marked boxes were determined
empirically using the CMD in Fig. \ref{f:cmdrad} and are designed to isolate the
``upper'' and ``lower'' turn-off stars. The main sequence itself has been excluded
to try and maintain as clean a separation as possible.}
\label{f:tosections}
\end{figure}

\begin{figure}
\centering
\includegraphics[width=0.5\textwidth]{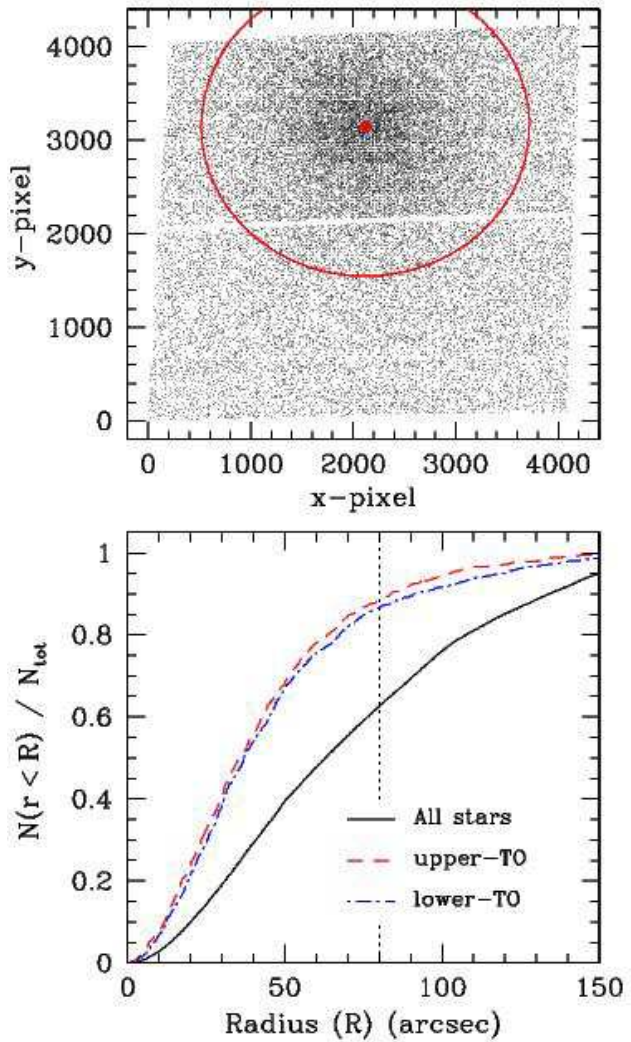}
\caption{{\bf Upper panel:} Pixel coordinates of all $34534$ detected stars with 
high quality photometry, along with our measured centre at $(x_c,y_c) = (2120,3150)$
(solid point). The solid circle marks a radius of $80\arcsec$ ($1600$ pixels)
about this centre. {\bf Lower panel:} Distributions of projected radial distance
from the cluster centre for all stars (solid line), upper turn-off stars (dashed
line), and lower turn-off stars (dot-dashed line). The vertical dotted line marks
a radius of $80\arcsec$, as plotted in the upper panel. Approximately $90$ per cent
of the turn-off stars lie within this radius.}
\label{f:toradial}
\end{figure}

We first need to demonstrate that all stars in the main sequence turn-off region
belong to the cluster. To this end, we used Fig. \ref{f:cmdrad} to isolate the
areas on the CMD covered by the two visible turn-offs. We then searched through
the full list of stars plotted in Fig. \ref{f:cmdall} and chose only those lying in
these two regions. The results of this process are displayed in Fig. 
\ref{f:tosections}. The two boxes define the sets of ``upper'' and ``lower'' turn-off
stars. By examining the spatial distributions of these two ensembles, we can
check that they both do belong to the cluster, and search for differences between
them.

In Fig. \ref{f:toradial} we plot the distribution of projected radial distance from 
the cluster centre for all stars and compare this with the distributions of projected
radial distance from the cluster centre for the upper and lower turn-off stars. We determined
the cluster centre using a surface-brightness testing algorithm very similar to that 
described by \citet{mackey:03}, with the result that the cluster centre lies within
$\approx 20$ pixels ($1\arcsec$) of the pixel coordinates $(x_c,y_c) = (2120,3150)$. The 
pixel coordinates of all stars are plotted in the upper panel of Fig. \ref{f:toradial} 
along with a point marking the measured cluster centre.

\begin{figure*}
\centering
\includegraphics[width=85mm]{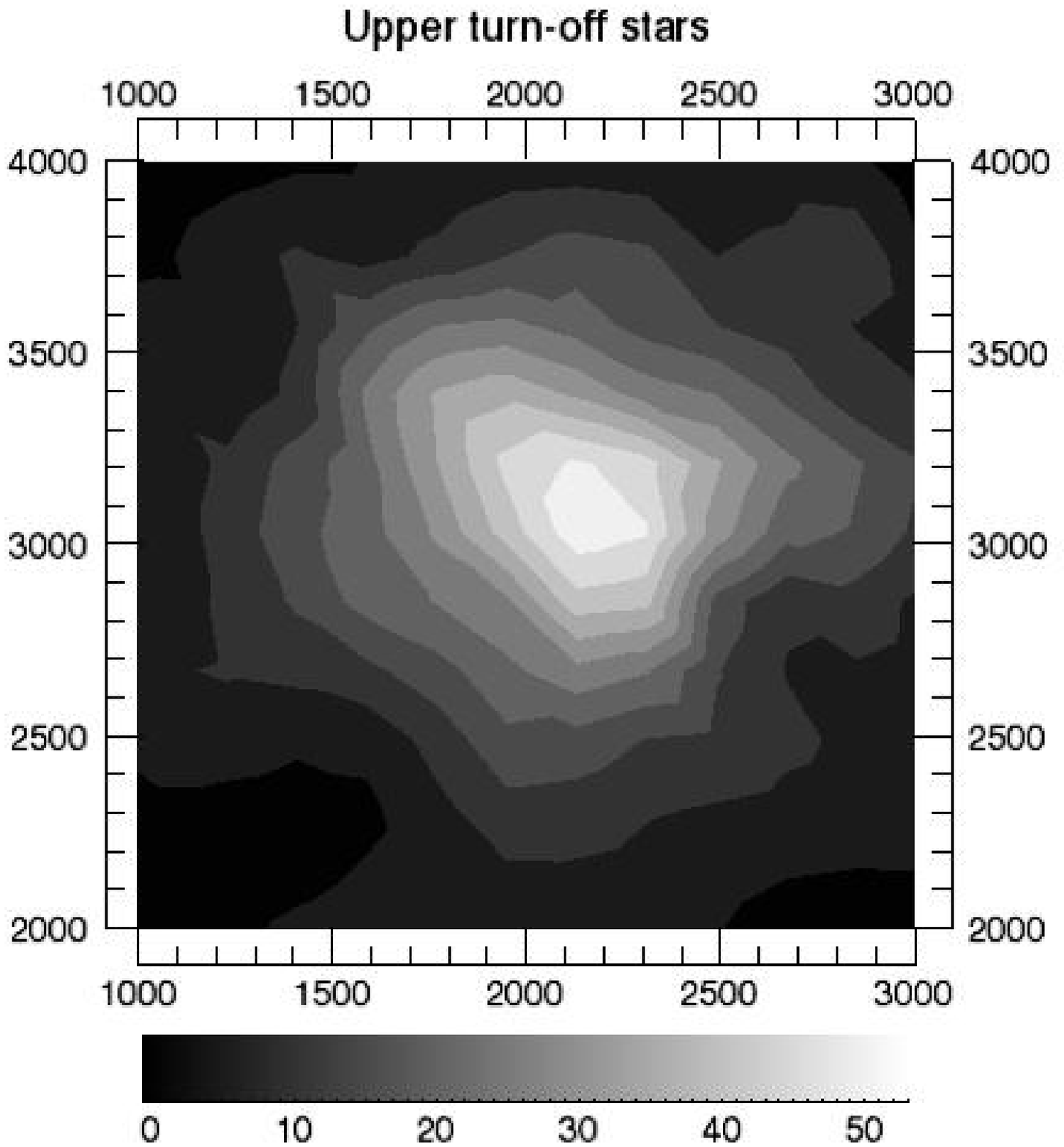}
\hspace{-2mm}
\includegraphics[width=85mm]{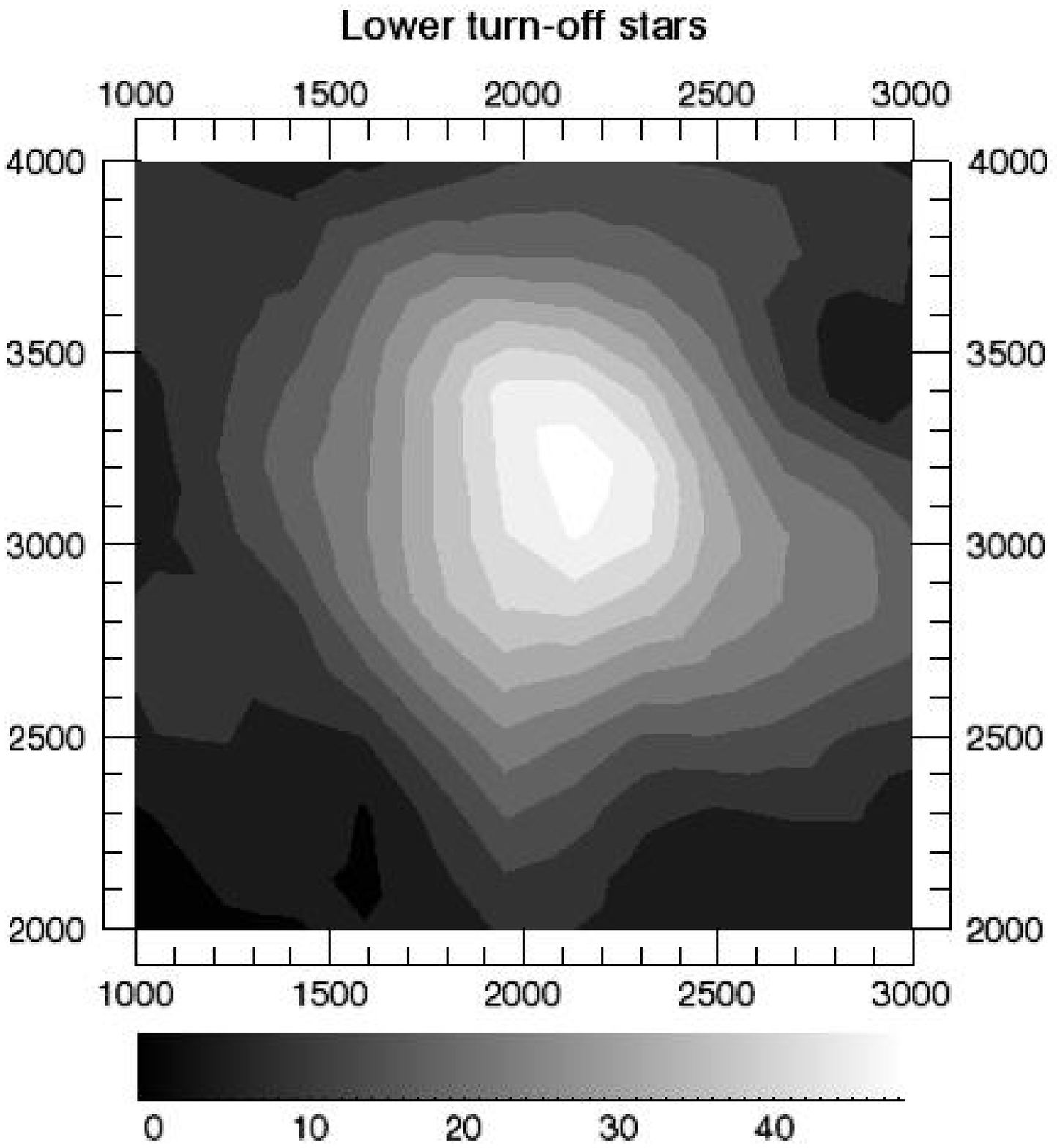} \\
\vspace{-3mm}
\includegraphics[width=85mm]{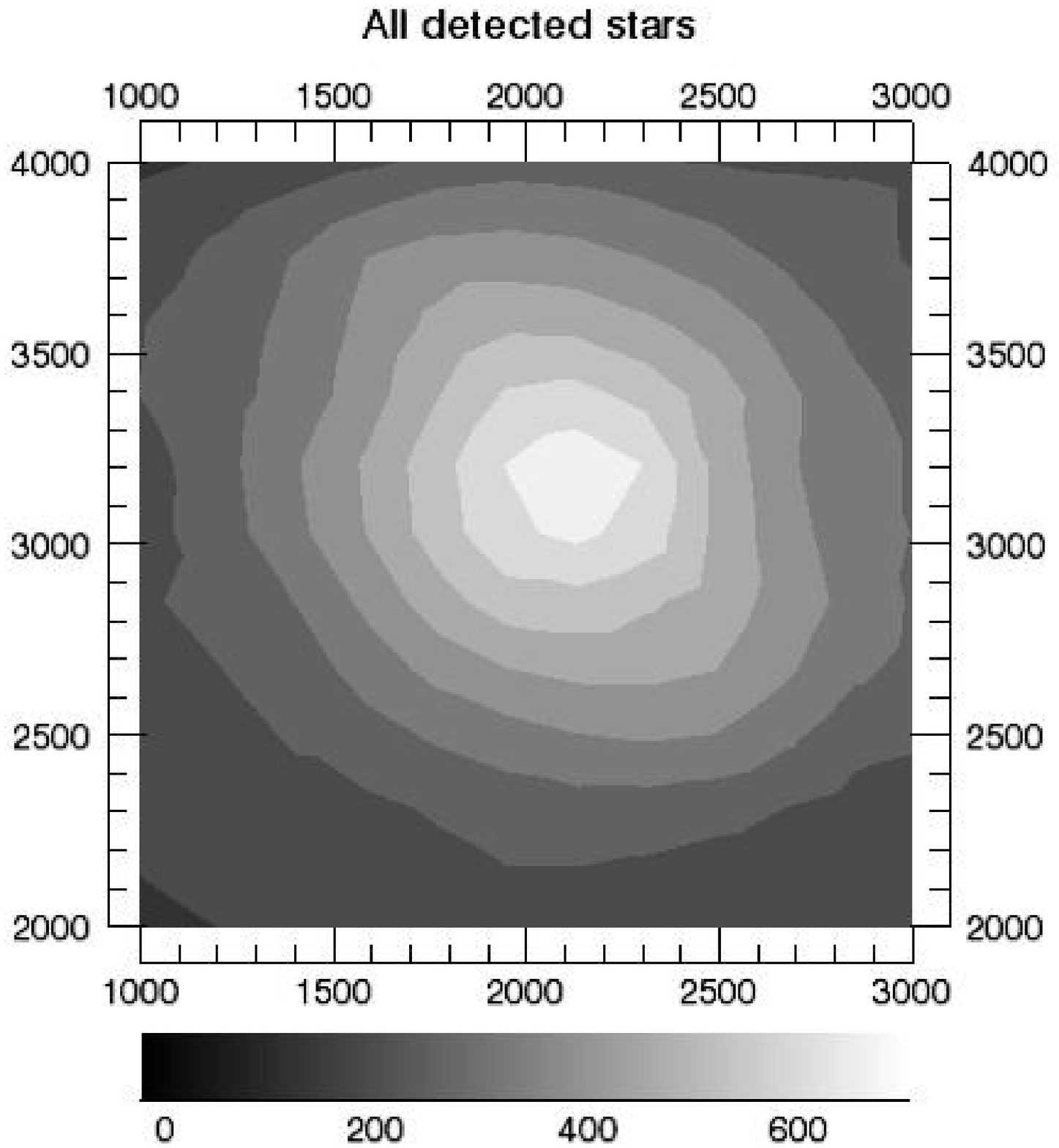}
\caption{Contour plots of the spatial distribution of upper turn-off stars (top left panel),
lower turn-off stars (top right panel), and all detected stars (lower panel). Axis units 
are pixel coordinates -- the plots are zoomed on the cluster centre, so do not cover 
the entire ACS field of view. They were produced by calculating a 2D histogram of stellar 
positions using a box size of $300\times300$ pixels, and then smoothing and contouring the 
result. The marked contours in the upper two panels delineate approximately equivalent 
density levels.}
\label{f:contours}
\end{figure*}

The radial distribution of all stars (including both cluster and field stars), 
plotted as a solid line in the lower panel of Fig. \ref{f:toradial}, is much less 
centrally concentrated than those of the upper and lower turn-off stars (dashed line, 
and dot-dashed line, respectively). Approximately $90$ per cent of the turn-off stars 
lie within a radius of $\sim 80\arcsec$ ($1600$ pixels) of the cluster centre, as marked 
in the upper panel of the Figure. This observation strongly implies that these objects 
are associated with the cluster itself rather than representing any type of field star 
contamination. There is also apparently a small difference between the radial 
distributions of the two sets of turn-off stars, suggesting that the lower turn-off 
stars may be a little less centrally concentrated than the upper turn-off stars.

In Fig. \ref{f:contours}, we show contour plots of the spatial distributions
of all stars (lower panel) and of both sets of turn-off stars (upper two panels). 
This allows us to further compare the spatial distributions of the turn-off stars 
to each other and to the rest of the cluster stars. We produced each plot by first
making a 2D histogram of the pixel coordinates of the stars in the relevant
group, using a bin size of $300\times300$ pixels. Selection of this size 
represents a trade-off between high spatial resolution and having a significant number
of stars per bin for the turn-off plots; we retained the same box size for 
the plot including all stars to allow direct comparison. After calculating the
histogram we smoothed the result using the {\sc min\_curve\_surf} function in
{\sc idl}, and then obtained a contour map using the {\sc contour} procedure
in {\sc idl}. The results still inevitably contain a small amount of residual 
boxiness from the 2D histogram; however this does not affect the overall shapes 
and spacing of the displayed contours.

The lower panel of Fig. \ref{f:contours} shows that the cluster considered as a 
whole appears to be relatively smoothly and symmetrically distributed. The centre 
matches well our measured value of $(x_c,y_c) = (2120,3150)$, and there is no strong
evidence for significant ellipticity or clumpiness. From the upper two panels, it
is clear that the two sets of turn-off stars are very definitely associated with the 
cluster, as we surmised from Fig. \ref{f:toradial}. In addition, within the accuracy 
of centroiding from the maps (roughly $\pm 50$ pixels, or $\pm 2.5\arcsec$) their 
centres are coincident with each other and the cluster as a whole. It is possible
that the set of upper turn-off stars is more centrally concentrated than both the 
set of lower turn-off stars and the cluster as a whole (this would be consistent
with Fig. \ref{f:toradial}), and furthermore, that the distribution of upper turn-off 
stars may also exhibit some significant asymmetry. However, both improved photometry 
of the turn-off region and a more sophisticated statistical analysis are required to 
draw a firm conclusion on these issues.

Together Figs. \ref{f:toradial} and \ref{f:contours} offer strong evidence that both 
sets of turn-off stars are definitely associated with NGC 1846 rather than any field 
star population. Furthermore, there are tantalizing indications that the spatial
distributions of the two sets of turn-off stars may be somewhat different to each
other, although improved measurements are required to confirm this.

\subsection{Isochrone fitting}
\label{ss:isochrone}
From the above analysis, we conclude that the observed double main sequence turn-off
in NGC 1846 results from the presence of two separate stellar populations in this
cluster. However, even though the two main sequence turn-offs are quite distinct, the RGB 
(below the saturation level at $m_{{\rm F555W}} = 18.0$) is narrow, and the RC is of a 
compact nature. This implies that the two populations cannot be of strongly differing 
metallicities, since both the RGB and the RC of a more metal rich population would lie 
considerably further to the red than those for a more metal poor population, and the RC 
would also be somewhat less luminous for the more metal rich population than for the
metal poor population. 

To quantify the main characteristics (age, metallicity) of the two populations in NGC 1846,
we fit isochrones to the cluster CMD. We used two sets of stellar models which
have isochrones calculated in the F555W and F814W ACS/WFC filter systems --
those of the Padova group \citep{girardi:00}, and those from the BaSTI evolutionary 
code \citep{pietrinferni:04,bedin:05}. We selected the ``basic set'' of Padova tracks 
with solar-scaled distribution of metals (in particular, that is, not enhanced in 
$\alpha$-elements), which include some degree of convective overshooting 
\citep[for details see][]{girardi:00}, and the ``non-canonical'' solar-scaled BaSTI 
tracks, which also employ convective overshooting \citep[see][]{pietrinferni:04}. 

Apart from the double turn-off, NGC 1846 presents a rather similar CMD to that
of NGC 1978, which was recently measured to be $1.9\pm0.1$ Gyr old 
\citep{mucciarelli:07}. There are two available spectroscopic measurements of the
metallicity of NGC 1846 -- that of \citet{olszewski:91} who found 
$[$Fe$/$H$] = -0.70\pm0.20$ from one RGB star, and that of \citet{grocholski:06} 
who obtained $[$Fe$/$H$] = -0.49\pm0.03$ from $17$ RGB stars.
For the Padova models we used the interactive web form available on the group's web 
site to construct a fine grid of isochrones about these values, sampling an age range
$1.0 \le \tau \le 3.0$ Gyr at intervals of $0.1$ Gyr, and a metal abundance range
$0.00250 \le Z \le 0.00950$ at intervals of $0.00025$. The total
metallicity $[$M$/$H$] = \log(Z/Z_\odot)$ where $Z_\odot \approx 0.019$,
so this abundance range corresponds to $-0.88 \le [$M$/$H$] \le -0.30$.
Assuming the $\alpha$-element enhancement in NGC 1846 is small, which would be
consistent with other intermediate-age LMC clusters 
\citep[see e.g.][ for NGC 1978]{mucciarelli:07}, then $[$M$/$H$] \sim [$Fe$/$H$]$.
For the BaSTI tracks, we used the web form available on that group's web site
to construct a grid of isochrones sampling the same age range as the Padova
grid, at the same intervals, but for only two metal abundances -- $Z = 0.004$
and $Z = 0.008$ (the BaSTI web form does not allow interpolation in metal
abundance).

We first attempted to find the best fitting isochrone for the upper main sequence 
turn-off. We did this by locating by eye three fiducial points on the CMD of the 
cluster's central region: the magnitude and colour of the turn-off, the magnitude 
of the RC, and the colour of the RGB at a level $1.0$ mag brighter than the level of 
the turn-off. This latter point was selected simply as a point lying on the lower
RGB at a level intermediate between that of the red end of the SGB and that of the 
RC. We then calculated the intervals $\Delta_{{\rm V}}$ and $\Delta_{{\rm C}}$, which 
are, respectively, the difference in magnitude between the level of the turn-off and 
the level of the RC, and the difference in colour between the turn-off and the RGB 
fiducial point. These intervals are useful because $\Delta_{{\rm V}}$ is strongly
sensitive to cluster age (and weakly to cluster metallicity), while $\Delta_{{\rm C}}$ 
is sensitive to both cluster age and metallicity. We obtained 
$\Delta_{{\rm V}} = 1.65 \pm 0.1$ mag and $\Delta_{{\rm C}} = 0.68 \pm 0.01$ mag. 

Next, we calculated the same intervals for all isochrones on each of our two grids, 
and selected only those with values lying within certain tolerances of the cluster
measurements ($\pm 0.2$ mag for $\Delta_{{\rm V}}$ and $\pm 0.02$ mag for 
$\Delta_{{\rm C}}$). This resulted in lists of fewer than ten isochrones per grid,
which we then fit to the CMD by eye. To do this, we calculated the offsets in 
magnitude and colour required to align the turn-off of the isochrone with that of
the CMD, and then those required to align the RC of the isochrone with that of the 
CMD. This resulted in two offsets in magnitude, and two offsets
in colour, which we averaged and then applied to overplot the 
isochrone on the CMD. Given the narrowness of most of the sequences on the CMD, it was 
then straightforward to identify the best fitting isochrone for each grid. The resulting 
offsets $\delta_{{\rm V}}$ and $\delta_{{\rm C}}$ provide estimates for the distance 
modulus to the cluster ($\mu$) and the foreground extinction, in that 
$\delta_{{\rm C}} = E(m_{{\rm F555W}}-m_{{\rm F814W}}) \approx E(V-I)$ 
\citep[since the extinction values are small, see][]{sirianni:05}
and $\delta_{{\rm V}} = \mu + 2.37 E(V-I)$.

We then moved to the lower turn-off and repeated the above process. In Section
\ref{ss:spatial}, we demonstrated that both populations in NGC 1846 share an
approximately common spatial centre. Furthermore, the narrowness of the CMD sequences 
implies that there is little or no differential reddening towards the cluster centre,
nor a significant line-of-sight depth to the system. In aligning the best-fitting
isochrones to the lower turn-off, we therefore applied the same distance
modulus and extinction values calculated for the upper turn-off.

\begin{figure}
\centering
\includegraphics[width=0.49\textwidth]{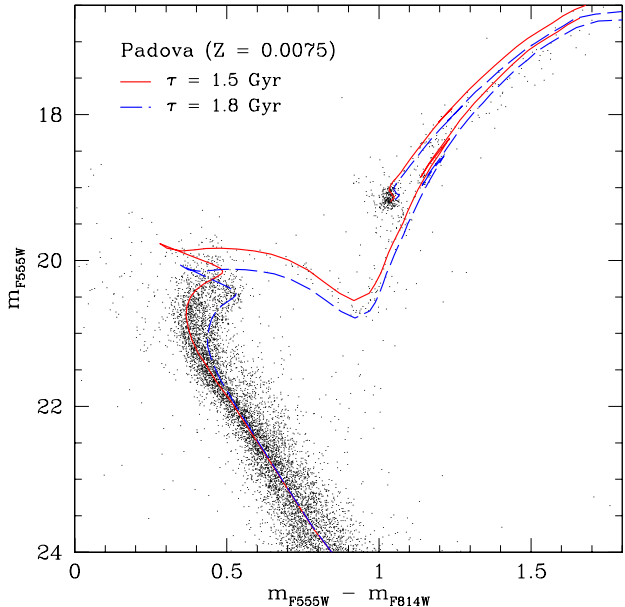}
\caption{Best-fitting Padova isochrones overplotted on the cluster CMD. The upper
and lower turn-off isochrones have the same metal abundance $Z = 0.0075$, and ages
$\tau_u = 1.5$ Gyr and $\tau_l = 1.8$ Gyr, respectively. Both
have been shifted by $E(V-I) = 0.045$ and $\mu = 18.37$ to achieve the best-fitting
alignment.}
\label{f:isopad}
\end{figure}

The best-fitting Padova isochrones may be seen in Fig. \ref{f:isopad}. We found 
that the highest quality solution was for isochrones of the same metal abundance
($Z=0.0075$, or $[$M$/$H$] = -0.40$) but differing ages: $\tau_u = 1.5$ Gyr for the 
upper turn-off and $\tau_l = 1.8$ Gyr for the lower turn-off. The isochrone
filtering process outlined above provided an indication of the random uncertainties
associated with these results -- isochrones with $\pm 0.00025$ in $Z$ and $\pm 0.1$
Gyr in age gave reasonable fits, but clearly not as good as for the selected values.
The fact that the two isochrones have the same metallicity is primarily driven
by the narrowness of the observed RGB and the compact nature of the RC. Matching
these features is successfully achieved if there is solely an age difference between 
the two populations. The observed width of the SGB is successfully 
matched at the red end, but is apparently a little too wide at the blue end.
The reason for this is not clear -- it may represent a shortcoming of the stellar
models (although we note that the Basti isochrones also predict too wide an SGB at 
the blue end), or may represent another very unusual feature of the CMD for NGC 1846.

The distance modulus and colour excess obtained for the fit described above are 
$\mu = 18.37$ and $E(V-I) = 0.045$. These are both slightly lower than the canonical 
values for the LMC; a similar effect was noted by \citet{mucciarelli:07} when fitting 
Padova isochrones to their CMD for NGC 1978.

\begin{figure}
\centering
\includegraphics[width=0.49\textwidth]{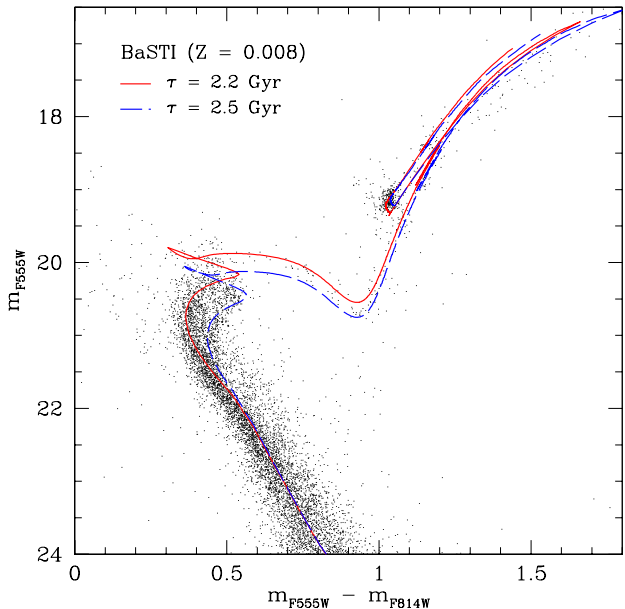}
\caption{Best-fitting BaSTI isochrones overplotted on the cluster CMD. The upper
and lower turn-off isochrones have the same metal abundance $Z = 0.008$, and ages
$\tau_u = 2.2$ Gyr and $\tau_l = 2.5$ Gyr, respectively. Both
have been shifted by $E(V-I) = 0.075$ and $\mu = 18.57$ to achieve the best-fitting
alignment.}
\label{f:isobasti}
\end{figure}

Fig. \ref{f:isobasti} shows the best-fitting BaSTI isochrones. For these models we
did not have the luxury of fine sampling in metal abundance; however, $Z=0.008$ 
is close to the best-fitting Padova metallicity, and provides adequate results. 
The best-fitting BaSTI isochrones are somewhat older than those
for the Padova models -- we found $\tau_u = 2.2$ Gyr for the upper turn-off and
$\tau_l = 2.5$ Gyr for the lower turn-off. Uncertainties are similar to those
for the Padova model fitting. The RGB, RC, and width of the SGB are
slightly more successfully matched than with the best-fitting Padova isochrones;
however, the turn-off regions are not quite as well matched. This may be an
artifact of having a slightly incorrect metallicity for the BaSTI models. The 
distance modulus and colour excess obtained for the fit are $\mu = 18.57$
and $E(V-I) = 0.075$. In this case, the distance modulus is slightly higher than
the canonical LMC value.

While our two pairs of measured absolute ages show an offset of $\sim 0.7$ Gyr, both 
sets of models show that the appearance of the NGC 1846 CMD can successfully be 
reproduced by adopting two stellar populations of the same metallicity and a difference 
in age of $\approx 300$ Myr. Our derived metallicity $[$M$/$H$] \approx -0.40$ is consistent
with the spectroscopic iron abundance measured recently by \citet{grocholski:06}
from a sample of $17$ RGB stars.

\section{Discussion and conclusions}
\label{s:discussion}
Our HST/ACS photometry of the rich intermediate-age LMC cluster NGC 1846 has revealed
an unusual CMD, exhibiting two main sequence turn-offs, but otherwise narrow and clearly
defined primary features (RGB, SGB, RC). By examining the spatial distributions of the
two sets of turn-off stars, we have shown that both are unequivocally associated with
the cluster rather than any field-star population. The set of upper turn-off stars
is possibly more centrally concentrated than that of the set of lower turn-off stars,
and may be asymmetrically distributed. Even so, the two sets of turn-off stars
share an approximately common centre. Via careful fitting of isochrones from two 
different groups of stellar evolution models (Padova and BaSTI), we have
demonstrated that the observed features on the CMD can be explained by assuming that
NGC 1846 harbours two stellar populations of the same metal abundance 
($[$M$/$H$] \approx -0.40$) but differing in age by $\approx 300$ Myr. The absolute
ages of the two populations are $\tau_u = 1.9 \pm 0.1$ Gyr and 
$\tau_l = 2.2 \pm 0.1$ Gyr. These are a straight average of the ages from the two 
different isochrone sets, and the uncertainties represent the
approximate random uncertainties associated with the fitting process. We note that
additional, larger systematic uncertainties are likely present -- the two pairs of
absolute ages are offset by $0.7$ Gyr between the two sets of isochrones.

Given these results, it is natural to ask how such an object as NGC 1846 might
have been formed. We can think of two options -- either this cluster underwent two
distinct episodes of star formation separated by $300$ Myr, or it is the result
of the merger of two star clusters formed $300$ Myr apart. 

Regarding the possibility that NGC 1846 underwent
two separate episodes of star formation, it is difficult to see how it can have
retained enough gas for this to happen. There is strong evidence that the combined
effects of massive stellar winds and supernova explosions expel any remaining
gas from a very young star cluster within only a few Myr 
\citep[e.g.,][]{bastian:06}, leading to strong changes in the cluster's internal
dynamics and structure \citep[see e.g.,][ and references therein]{goodwin:06}.
A cluster must be very massive to retain any gas in its potential well. The 
only clusters where multiple episodes of star formation appear evident are
globular clusters at the very upper end of the globular cluster mass function,
such as $\omega$ Centauri \citep[e.g.,][]{villanova:07}, and G1 in M31 \citep{meylan:01}. 
NGC 1846 does not fulfil this criterion. Furthermore, in all such cases where 
multiple episodes of star formation have taken place, self-enrichment has apparently 
occurred. The timescale for such enrichment is not well defined; however, it seems 
possible that $300$ Myr could be long enough given that significant chemical processing 
must have already occurred in the most massive cluster stars (in particular in type II 
supernovae and massive asymptotic giant branch stars), within the first
$\sim 100$ Myr. It is therefore not clear why the two populations in NGC 1846 
should apparently have the same metal abundance.

The merger scenario fares better. It is well known that both the LMC
and SMC possess populations of candidate binary (and multiple) star clusters 
\citep[e.g.,][]{bhatia:88,hatzidimitriou:90}, and by using statistical
arguments it can be shown that a significant fraction of these are likely to
be physically linked \citep{bhatia:88}. Gravitationally bound star clusters
are predicted to merge on time-scales of tens of Myr to a few hundred Myr 
\citep[e.g.,][]{spz:06}, which is roughly consistent with the observed age
difference between the two populations in NGC 1846. Old binary pairs -- such 
as SL 349-SL 353 with an age $\sim 500$ Myr -- are certainly observed in the LMC
\citep*{leon:99}. However, theories for the formation of bound binary star clusters 
\citep[e.g.,][]{fujimoto:97} suggest that such objects should be roughly coeval, 
with age differences less than $\approx 100$ Myr at most. This is inconsistent 
with the observed populations in NGC 1846, and indeed with several known binary 
pairs in the LMC -- such as SL 356-SL 357, which have ages of $70$ Myr and 
$600$ Myr, respectively \citep{leon:99}. 

One way to circumvent this is if some clusters are formed in star cluster groups 
(SCGs) in giant molecular clouds \citep[see e.g.,][ and references therein]{leon:99}. 
In such complexes, cluster formation can be spread over several hundred Myr. Furthermore, 
clusters observed to be physically associated at present do not have to have been 
formed this way, as the cross-section for tidal capture is much increased due to 
the relatively high density of star clusters within the SCG. This scenario therefore
allows for the existence of cluster pairs which are older than the typical 
time-scale for merging, as well as for cluster pairs with large age differences,
both of which types are seen in the LMC. If the two components of NGC 1846 were 
formed in a SCG in the same giant molecular cloud, this also naturally explains 
their matching metal abundances.

In this scenario, the two clusters comprising the present day NGC 1846 must have
merged less than $1.5-2.2$ Gyr ago, based on our derived ages for the upper
turn-off population. The median relaxation time in intermediate-age LMC clusters
is typically of order $1-2$ Gyr \citep[e.g.,][]{mackey:07}; the central relaxation 
time may be up to a factor ten shorter than this. The very inner region of 
NGC 1846 is therefore quite possibly dynamically old enough for two merging 
clusters to now be sharing a common centre and be fairly well mixed through
relaxation processes; however it is probable that asymmetries could still
exist outside the core, as our observations have hinted. It would be extremely 
interesting to investigate the evolution of a merged LMC-type cluster via realistic
$N$-body modelling to examine if the observed features of NGC 1846 can be
reproduced. Such calculations may allow the time when merging occurred to be
constrained -- this would be extremely useful in the context of examining the
SCG scenario. Direct $N$-body modelling of LMC-type clusters is now possible
-- \citet{mackey:07} present models with $N=10^5$ particles integrated over
a Hubble time of evolution. Since we do observe asymmetries in the distribution
of the two populations in NGC 1846, dynamical signatures offering clues to the
formation process may also still be present outside the cluster centre. It would 
certainly be interesting (although observationally challenging) to investigate the 
internal dynamics of this object. $N$-body modelling would be helpful in this regard 
as it would allow the expected level of the signature to be predicted. NGC 1846 
patently warrants further study, including improved photometry of the turn-off 
region, for important insights into star cluster formation processes.

\section*{Acknowledgements}
We are grateful to Annette Ferguson for reading through a draft of this paper
and offering helpful comments and suggestions. ADM is supported by a Marie Curie 
Excellence Grant from the European Commission under contract MCEXT-CT-2005-025869.
This paper is based on observations made with the NASA/ESA Hubble Space Telescope, 
obtained at the Space Telescope Science Institute, which is operated by the Association of 
Universities for Research in Astronomy, Inc., under NASA contract NAS 5-26555. 
These observations are associated with program \#9891.

\bsp

\label{lastpage}

\end{document}